\documentclass[12pt]{iopart}

\usepackage{graphicx}
\usepackage{cite}
\usepackage{subfig}
\begin{document}

\title[Crust EOS and GW170817 analysis]{The impact of the crust equation of state on the analysis of GW170817} 

\author{R. Gamba$^{1,2}$, J. S. Read$^3$, L. E. Wade$^4$} 

\address{$^1$ INFN Sezione di Torino, Via P. Giuria 1, 10125 Torino, IT}
\address{$^2$ Universit\`a di Torino, Via P. Giuria 1, 10125 Torino, IT}
\address{$^3$ GWPAC, California State University Fullerton, Fullerton, CA 92831, USA}
\address{$^4$ Kenyon College, Gambier, OH 43022, USA}

\ead{rossella.gamba@edu.unito.it}

\begin{abstract}
The detection of GW170817, the first neutron star-neutron star merger observed by Advanced LIGO and Virgo, and its following analyses represent the first contributions of gravitational wave data to understanding dense matter. Parameterizing the high density section of the equation of state of both neutron stars through spectral decomposition, and imposing a lower limit on the maximum mass value, led to an estimate of the stars' radii of $R_1 = 11.9_{- 1.4}^{+ 1.4}$ km and $R_2 = 11.9_{- 1.4}^{+ 1.4}$ km \cite{Abbott:2018exr}. These values do not, however, take into account any uncertainty owed to the choice of the crust low-density equation of state, which was fixed to reproduce the SLy equation of state model \cite{Douchin:2001sv}. We here re-analyze GW170817 data and establish that different crust models do not strongly impact the mass or tidal deformability of a neutron star -- it is impossible to distinguish between low-density models with gravitational wave analysis. However, the crust does have an effect on inferred radius. We predict the systematic error due to this effect using neutron star structure equations, and compare the prediction to results from full parameter estimation runs. For GW170817, this systematic error affects the radius estimate by 0.3 km, approximately $3\%$ of the neutron stars' radii.
\end{abstract}
\maketitle

\section{Introduction}

The composition and structure of neutron stars (NS) is a longstanding question for discussion in the scientific community. Knowing the properties of these very dense objects would contribute to the understanding of matter under extreme conditions, with implications for both astrophysics and nuclear physics, as reviewed in e.g. \cite{Lattimer/annurev-nucl-102711-095018, Hebeler:2015hla, Özel/annurev-astro-081915-023322, Miller:2016pom}.

In August 2017, the first observation of a NS-NS merger, GW170817 \cite{TheLIGOScientific:2017qsa}, was made by the Advanced LIGO \cite{TheLIGOScientific:2014jea} and Virgo \cite{TheVirgo:2014hva} instruments. 
Its subsequent analysis represents the first contribution of gravitational-wave (GW) data to understanding dense matter and the NS radius $R$ \cite{TheLIGOScientific:2017qsa, Abbott:2018exr, Annala:2017llu, De:2018uhw, Fattoyev:2017jql, Most:2018hfd, Raithel:2018ncd, Tews:2018chv}. Rapid detection and localization of the signal \cite{Veitch.PRD,Singer.PRD,Messick:2016aqy, Nitz:2017svb } enabled multimessenger followups \cite{GBM:2017lvd,Monitor:2017mdv, Coulter:2017wya, Goldstein:2017mmi, Haggard:2017qne, Hallinan:2017woc, Troja:2017nqp},
which themselves suggest various scenarios that may also limit the properties of cold dense matter, e.g. \cite{Margalit:2017dij, Nicholl:2017ahq, Radice:2017lry, Shibata:2017xdx}.  We here focus on what GW data alone implies.

Comparing GW data to solutions of the relativistic two-body problem provides a measurement of the two stars' masses $M_1$ and $M_2$ and dimensionless tidal deformabilities $\Lambda_1$ and $\Lambda_2$, which describe the ratio of the body's tidally induced quadrupolar deformation to the tidal potential caused by its companion, through the mass-weighted sum $\tilde{\Lambda}$. 
The tidal parameters depend on the compactness of the star $C = R/M$, both explicitly and through the relativistic tidal Love number $k_2$ \cite{TDfirst, PhysRevD.43.3273, PhysRevD.45.1017, PhysRevD.47.3124, PhysRevD.49.618, Flanagan:2007ix, Hinderer:2007mb,Damour:2009vw}. 

When determining $R$ from GW analysis, an added level of uncertainty comes from the choice of the crust structure model. The outer low-density layers of the star contain a small fraction of the mass (for a
M = 1.4 $ M_{\odot}$ NS with a SLy equation of state, $M_{crust}/M \approx 1\%$ below $\rho_{crust} \approx 1.4 \times 10^{14} g/cm^3$), but contributes a larger portion of the radius ($R_{crust}/R \approx 6\%$).
If the tidal deformability parameters $\Lambda$ are mostly unaffected by the choice of the crust, then it will not be possible to distinguish  between different low-density models through GW analysis, de-facto adding a ``systematic error'' on the values of the NS radii thus obtained. Notably, one implication of the low sensitivity of tidal parameters to the crust densities would be that GW measurements give more direct information on higher densities, and that therefore in this region the constraints obtained from analyses are independent of uncertainties in crust. 
 
The aim of this paper is to quantify the effect of the choice of the crust equation of state (EOS) on parameter estimation (PE) for GW170817, especially on the radius estimate of \cite{Abbott:2018exr}. In order to do so, we re-analyse the GW data with \verb|LALInference| 
\cite{lalsuite}, following the parameterized EOS method of \cite{Carney:2018sdv}, but
modifying the low-density region. In the LIGO-Virgo analysis \cite{Abbott:2018exr}, densities below $\rho \approx 10^{14} g/cm^3$ were fixed to the SLy description of \cite{Douchin:2001sv}.
We replace the fixed region with crust EOSs described in \cite{Newton:2011dw}, which were obtained through the combination of  Compressible Liquid Drop Model (CLDM) \cite{Baym:1971pw} and Baym, Pethick and Sunderland (BPS) models \cite{Adam:2015ele}, while continuing to parameterize the higher-density core with a spectral decomposition following \cite{Lindblom:2010bb}. In parallel, we try to predict the effect of this change without full reanalysis: by gluing different crusts to core EOSs recovered in \cite{Abbott:2018exr},  we obtain a quantitative prediction of the impact of the crust on the radii recovered for the NSs involved in the GW170817 coalescence.
We find that varying the crust has negligible impact on the $\Lambda$ and M distributions, but can shift the implied radius up by $\approx 0.3$ km in the full PE.

The paper is organized as follows: section 2 is dedicated to the NSs' EOS, focusing on its parametrization in the core of the star and on choosing appropriate crust models with respect to the existing bounds on symmetry energy and its slope, two important nuclear parameters whose meaning will be briefly presented;
section 3 describes the methods used to estimate and predict the stellar parameters' values; finally, in sections 4 and 5 the estimate of the systematic error entailed by ignoring crust variations is computed and discussed.

\section{Equation of State}
Statements on the behaviour of NS matter quantitatively translate into imposing constraints on its energy density - pressure relationship, the EOS, which then lead to limits on stellar parameters such as maximum mass and radius. In this section we first give a quick overview of the crust composition and of the model EOS chosen to describe it. We then introduce the spectral decomposition parametrization of the high-density EOS adopted in our \verb|LALInference| run.

\subsection{Crust Equation of State}
NSs are objects so dense that it is possible for them to develop a sturdy crust made of very neutron-rich nuclei even at the incredibly high temperatures of their surface ($T \approx 10^5-10^6 K$)\cite{Chamel:2008ca}. Qualitatively, one could imagine that when moving farther away from the core of the star - as the temperature and density decrease - the Coulomb interactions between particles become more and more important with respect to their thermal and quantum energy. At one point, this translates into the formation of neutron-rich nuclei, their locking from nucleon plasma into a lattice and the creation of a solid layer.
In the innermost part of this solid crust, still very close to the core, the extreme density conditions may cause the lattice nuclei to change shape: no longer spherical, the minimum-energy structures could vaguely resemble pasta forms. At lower densities the nuclei go back to their spherical form, but are still so neutron-rich that some neutrons ``drip out'' of them and, if the temperature is below a critical value, form a superfluid neutron vapour. Approaching the exterior of the crust, the drip phenomenon stops; this transition divides the inner from the outer crust, which is characterized by the presence of a heavy nuclei lattice immersed in an electron gas.
To quantitatively describe this complex behaviour, a number of models have been developed for both the inner (Thomas-Fermi, CLDM) and outer crust (BPS) \cite{Baym:1971pw, Oyamatsu:1993zz, Chamel:2008ca, Adam:2015ele}. 

In general, whatever combination of models is chosen to describe the complete low-density section can be characterized by two parameters defined at nuclear saturation density $n_0$: the symmetry energy $S_0$, which encodes the energy cost of making NS matter more neutron-rich, and its slope $L$.
Following common notations and defining the fraction of neutrons to all baryonic matter as $x$ and the isospin asymmetry  as $\delta = 1 - 2x$, through a Taylor expansion around $x = 1/2$ of the energy density per nucleon $E$ one finds: 
\begin{equation}
 E(n, x) = E_0(n, 1/2) + S(n, 1/2)\delta^2 + \dots
\end{equation}
with 
\begin{equation}
S(n, 1/2) = S_0 + L (n-n_0)/3 n_0  + \dots
\end{equation}
The slope parameter $L$ plays an exceptionally important role in NS structure, as it is closely related to the pressure of purely neutron matter at sub-saturation densities through $ p(n, 1) = n^2/(3n_0) (L + \dots) $.

\begin{figure}
\begin{minipage}[b]{0.45\textwidth}
\centering
\includegraphics[width=\textwidth]{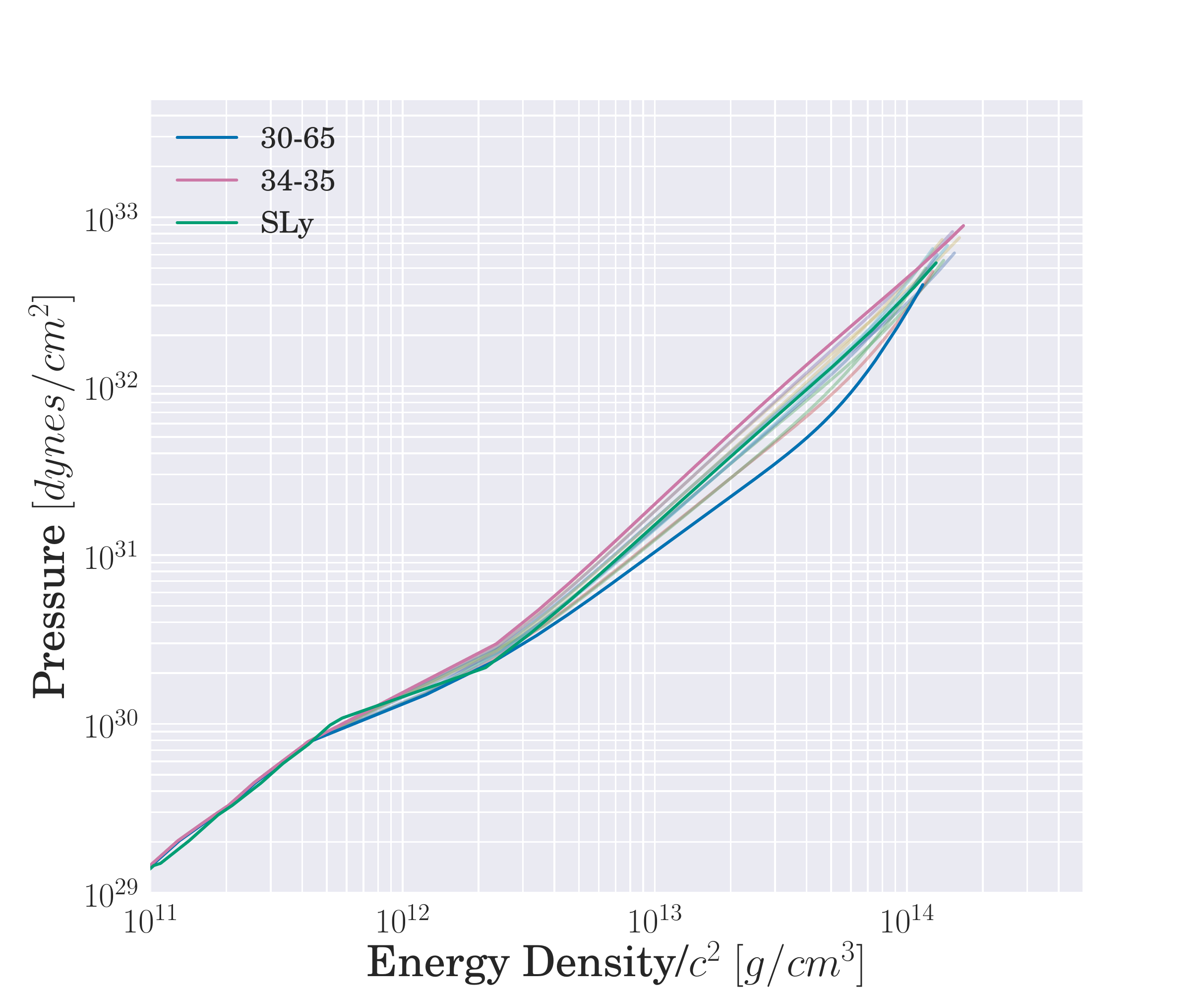}
\caption{\label{crust_eos} The EOSs for realistic crust models with nuclear parameters $S_0 \in [30-34]$ MeV and $L \in  [30-70] $ MeV. SLy is characterized by $S_0 = 32$ MeV and $L = 46$ MeV. The 34-35 MeV and the 30-65 MeV EOSs are the upper and lower limits.} 
\end{minipage}
\hfill
\begin{minipage}[b]{0.45\textwidth}
\centering
\includegraphics[width=0.91\textwidth]{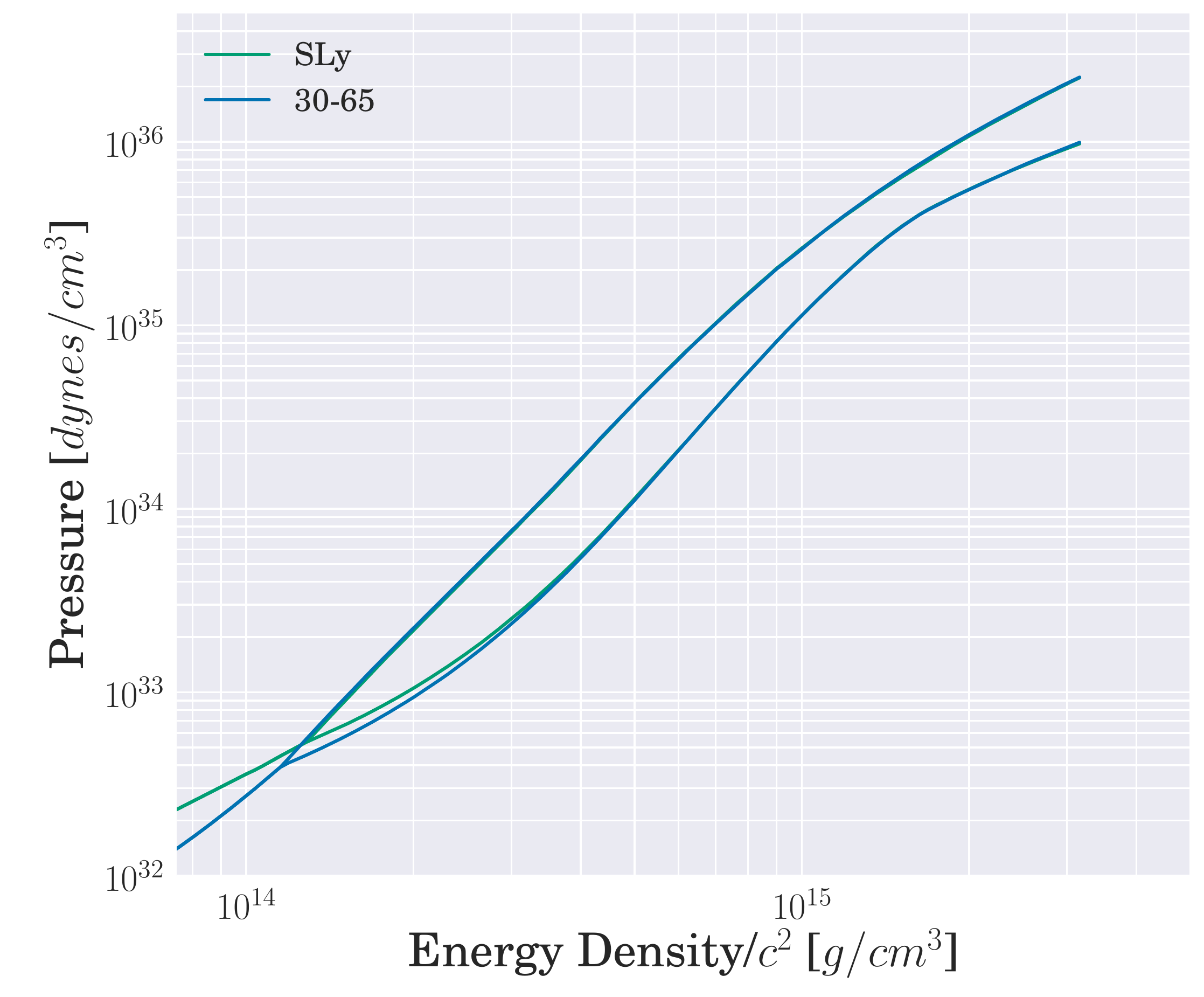}
\caption{\label{cls}90\% pressure-density credible levels. They are perfectly superimposed, except for the point where they are matched to the crust: credible levels on the high density section of the EOS are independent on the choice of the crust.}
\end{minipage}
\end{figure}

From \cite{Newton:2011dw} we retrieve a set of crust EOSs computed through a CLDM+BPS model, which reach up to approximately $\rho_{crust} = 10^{14} g/cm^3$ and cover a wide region of the $S_0 - L$ plane. Both $S_0$ and $L$ have been studied and constrained by a number of independent terrestrial experiments, which performed measurements of giant dipole resonances and dipole polarizabilities, nuclear masses, flows in heavy-ion collisions and neutron-skin thicknesses \cite{Lattimer:2012xj, Lattimer:2014sga, Danielewicz:2016bgb}. Taking into account all constraints, $S_0$ should range from 30 to 32 MeV and $L$ from 40 to 60 MeV for $L$ \cite{Kolomeitsev:2016sjl}.
Since, however, the acceptable ranges of $S_0$ and $L$ are still uncertain, we focus on the larger intervals 30 to 34 MeV for $S_0$ and 30 to 70 MeV for $L$ \cite{Lattimer:2015nhk}.
We then select the upper and lower limits EOS curves, whose parameters are respectively $S_0 = 34$ MeV, $L = 35$ MeV  and $S_0 = 30$ MeV, $L = 65$ MeV. These should give the largest impact on neutron star structure (see figure \ref{crust_eos}). We note that, although we use variation of $S_0$ and $L$ to establish a realistic range of crusts, we do not enforce an extrapolation to higher densities that is consistent with the chosen parameters, but allow the core EOS to vary independently of the chosen crust.

\subsection{Parametrization model}

There are two necessary conditions that a NS EOS has to satisfy to be physically consistent: sound has to propagate through the star slower than light in vacuum (causality) and pressure $p$ must be a monotonically increasing function of energy density $e$. When parametrizing an EOS, it would then be particularly convenient to choose a model that automatically fulfils at least the second condition. Such a parametrisation could be obtained through piece-wise polytropes \cite{Read:2008iy,Lackey:2014fwa} or by applying a spectral decomposition \cite{Lindblom:2010bb} on a basis of differentiable functions. Additionally, when fitting a known EOS, spectral fits frequently have smaller residuals than piecewise-polytrope fits, even when they are performed employing fewer parameters than piecewise-polytrope fits. 

The basic idea behind the spectral decomposition of \cite{Lindblom:2010bb} is that of expressing the adiabatic index $\Gamma(p) = [(e + p)/p] dp/de$, unique up to an integration constant for each EOS,  as the exponential of the sum of some smooth basis functions - $f^i = [ln(p/p_0)]^i, i \in \mathbf{N}$ in our specific case - multiplied by some coefficients $\gamma_i$:

\begin{equation}
 \Gamma(p) = exp[\sum_{i = 0}^{\infty}{\gamma_i f^i(p)}]
\end{equation}

Obtaining the expression of the energy density as a function of pressure requires then a simple integration:
\begin{equation}
e(p) = \frac{e_0}{\mu(p)} + \frac{1}{\mu(p)} \int_{p_0}^{p}{\frac{\mu(p')}{\Gamma(p')} dp'}
\label{e_p_spectral}
\end{equation}
in which $e_0$, $p_0$ is the starting point of the decomposition in the energy density-pressure plane and
\begin{equation}
\mu(p) = exp[-\int_{p_0}^{p}{\frac{dp'}{p'\Gamma(p')}}]
\end{equation}

\section{Parameter Estimation}

Running PE on the real data of GW170817 using spectral decomposition means that every spectral coefficient $\gamma_i, i \in [0,3]$ is sampled, in place of the tidal deformabilities $\Lambda_1$ and $\Lambda_2$, by stochastically walking through the parameter space \cite{Carney:2018sdv}. In order to cover a wide range of candidate EOSs, we sample $\gamma_0 \in [0.2, 2]$, $\gamma_1 \in [-1.6, 1.7]$, $\gamma_2 \in [-0.6, 0.6]$ and $\gamma_3 \in [-0.02, 0.02]$. We additionally impose that the adiabatic index $\Gamma(p) \in [0.6,4.5]$, as done in \cite{Abbott:2018exr}.
Each set $\{\gamma_0,\dots, \gamma_3\}_i$ can be mapped through (\ref{e_p_spectral})  in an EOS $p_i(e)$, $i = 1, \dots, N_{samples}$. The pressure-density credible levels can be obtained by choosing a list of $N'$ energy density values ${e_j}$, $j = 1, \dots, N'$, creating a pressure histogram for each value on the list by evaluating $p_i(e_j)$  $\forall i$, and finally finding the chosen percentiles from every histogram. 
To then go from an EOS of the form $e(p)$ to the determination of the stellar parameters $M$, $R$ and $\Lambda$, one has to integrate TOV equations and solve the inner-outer matching problem related to the relativistic Love number $k_2$. 
We used the publicly available code of \verb|LALSimulation|\cite{lalsuite} for such operations.
By additionally imposing a lower limit of 1.97 solar masses on the maximum mass value supported by the EOS, and fixing the low-density section of the EOS to reproduce the SLy model, the radii of the two NSs involved in the binary coalescence of GW170817 were estimated to be $R_1 = 11.9^{+1.4}_{-1.4} $ km and $R_2 = 11.9^{+1.4}_{-1.4}$ km.
In our \verb|LALInference| Markov chain Monte Carlo (MCMC) run, the hard-coded SLy crust was switched to the $S_0 = 30 $ MeV $L = 65$ MeV EOS (figure \ref{crust_eos}), and the crust-core transition point $e_t, p_t$ changed accordingly. The approximant used, IMRPhenomPNRTidal \cite{PhysRevD.86.104063, Hannam:2013oca,Schmidt:2014iyl, Husa:2015iqa, Khan:2015jqa}, and the choice of the other priors match the settings outlined in \cite{Abbott:2018exr}. 
The radii found then are $R_1 =11.7_{- 1.4}^{+ 1.4}$ km, $R_2 =11.7_{- 1.4}^{+ 1.3}$ km.
Pressure-density credible levels have been computed, and are shown in figure \ref{cls}.

While full MCMC runs can paint a precise picture of what happens when changing the low-density EOS, they also are very computationally-expensive and time demanding. For this reason, it is worth trying to make some rough - but fast - predictions, which would also allow one to check on the possible impact of some future variations, with a better idea of how well it will reflect the full analysis. 
Indeed, working under the assumption that both the masses and the core EOS are weakly affected by the choice of the crust, we can use the mass and spectral coefficients posteriors from \cite{Abbott:2018exr} to predict the posterior distributions of the stellar parameters $R$ and $\Lambda$ that we would get with our new crust.
We replace the SLy EOS with the 30-65 (34-35) crust and glue the final points of the new EOS to the final point of original the SLy crust table ($e_{m}, p_{m}$). Then, for pressures $p$ and energy densities $e$ higher than $p_m$ and $e_m$ we compute $e(p)$ through (\ref{e_p_spectral}), using as $\{\gamma_i\}$ those obtained from the posterior distribution of the previous analysis. Coupling these relations to the mass posteriors, we can compute the predicted distributions of the radii $R^{pr}$ of the NSs and of their tidal deformabilities $\Lambda$ (figures \ref{2D} (a) and (b)).
The radii values found are $R_1^{30-65 pr} = 11.7_{- 1.3}^{+ 1.4}$ km and   $R_2^{30-65 pr} = 11.7_{-1.3}^{+1.4}$ km;  $R_1^{34-35 pr} = 12.0_{- 1.4}^{+ 1.5}$ km and  $R_2^{34-35 pr} = 12.0_{- 1.4}^{+1.5}$ km.

We again note that at densities from approximately $10^{14} g/cm^3$ and up the EOS curves obtained through spectral decomposition do not necessarily have the characteristics at saturation density ($\rho_0 = 2.8 \times 10^{14} g/cm^3$) implied by the lower-density crust. Our aim, instead, is to compare directly with the results of \cite{Abbott:2018exr}, and estimate the effect that changing only the previous hard-coded outer crust has on PE.
If we were to impose consistency on $S_0$ and $L$ when sampling the spectral coefficients, or if we extended the fixed EOS region from crust through to $\rho_0$ with the same $S_0$ and $L$, we expect that we would find increased correlation between crust and radius results as was seen in \cite{Fortin:2016hny}. 
Such correlations would come through $S_0$ and $L$ choices rather than from the crust densities themselves.

\section{Systematic Error Estimate}

\begin{figure}[h]
\centering
\subfloat[][]
{\includegraphics[width=.45\columnwidth]{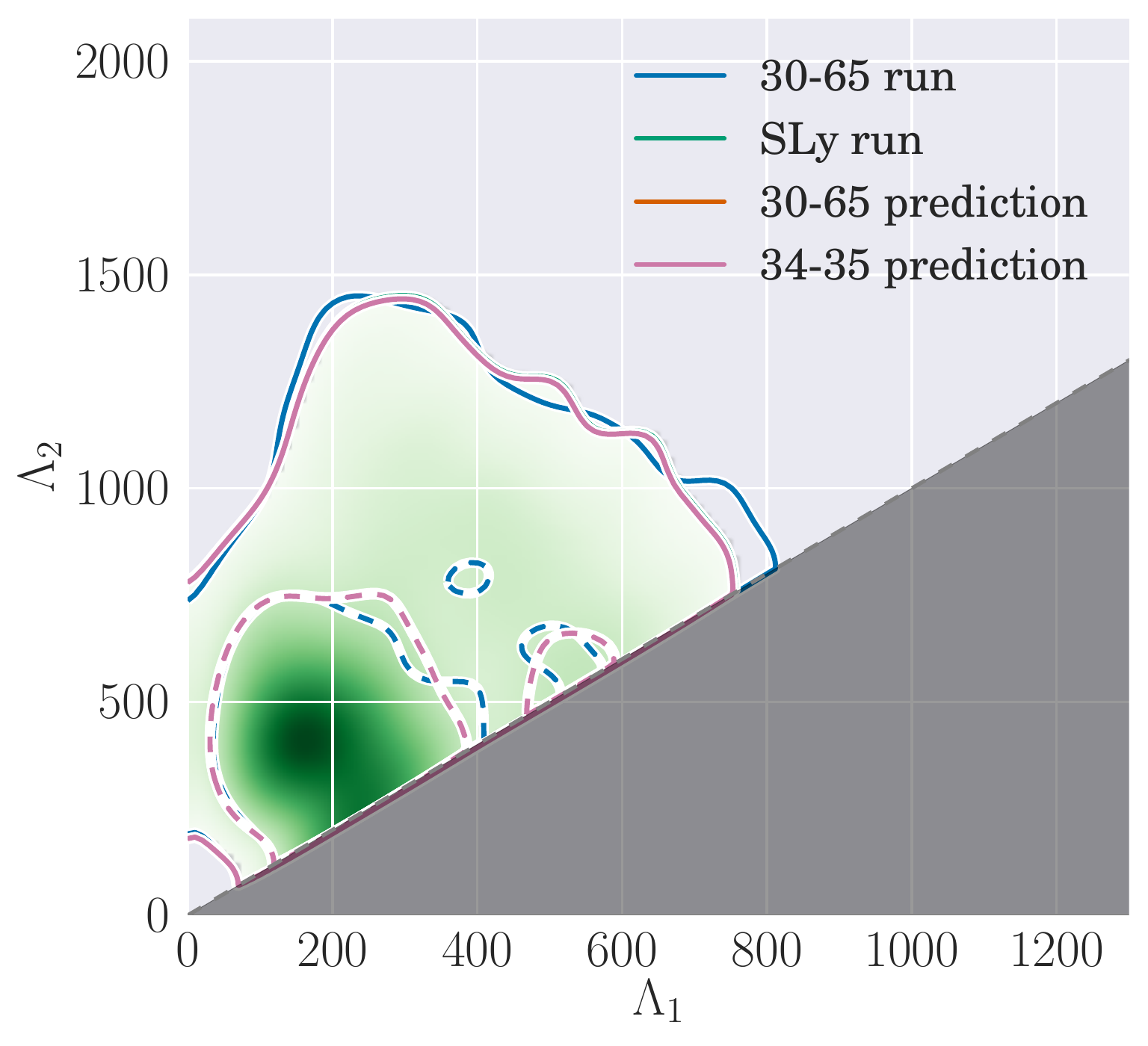}} \quad
\subfloat[][]
{\includegraphics[width=.45\columnwidth]{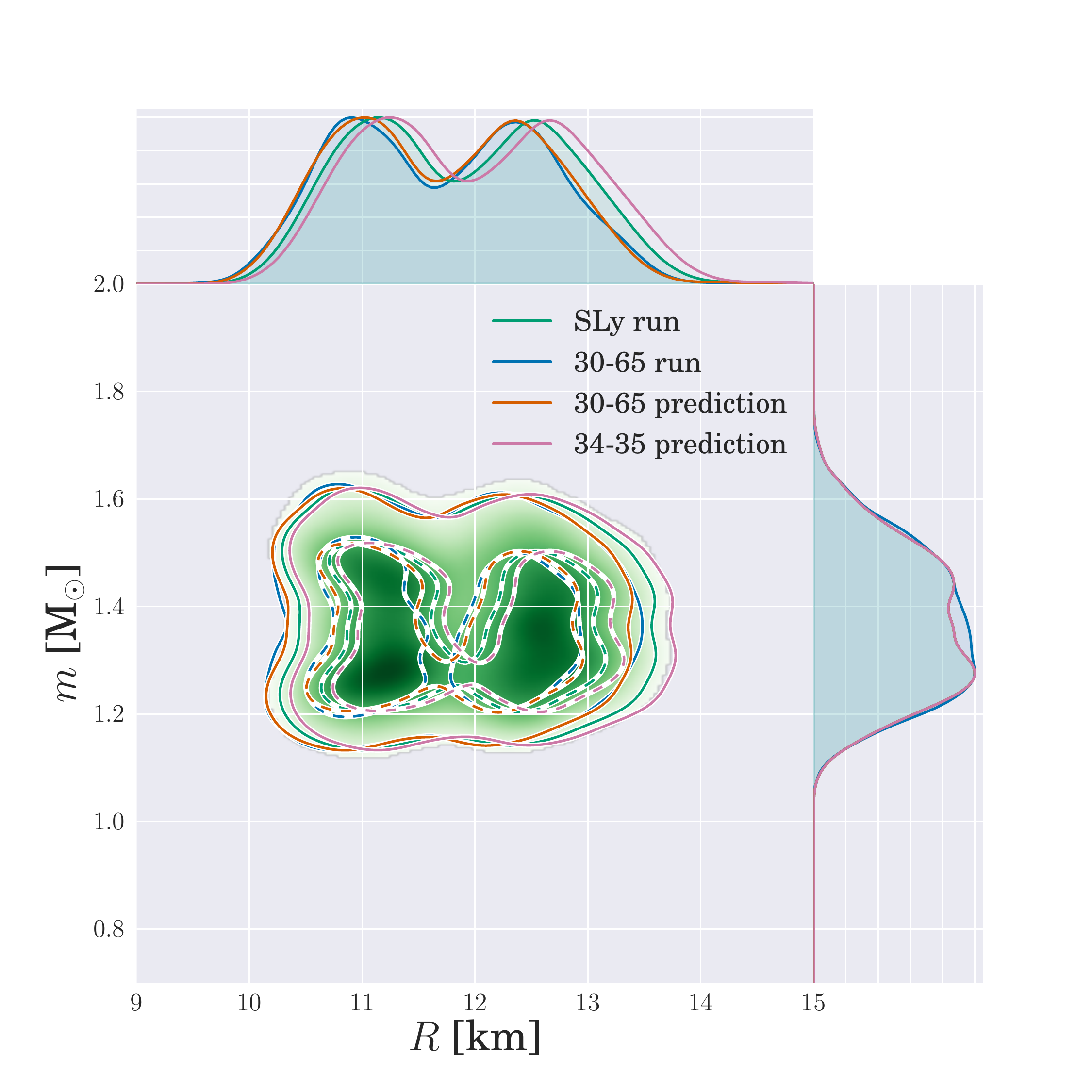}} \\
\caption{The $\Lambda_1$ vs $\Lambda_2$ (a) and mass vs radius (b) 2D distributions. The continuous and dashed curves represent, respectively, the 90\% and 50\% credible limits. While the $\Lambda$ distributions are all indistinguishable - and the ``prediction'' curves perfectly superimposed! - the radii distributions obtained with the 30-65 and 34-35 crusts are systematically shifted with respect to each other and to the one resulting from the SLy low-density model. This shift measures the additional uncertainty, in radius only, due to the unknown crust EOS. The GW constraints on tidal deformation are insensitive to the crust.}
\label{2D}
\end{figure}

In figure \ref{2D} (a)
we plot the 2D $\Lambda_1$ vs $\Lambda_2$ distributions, obtained through \verb|LALInference| runs and through the predictions we made. They are all almost perfectly superimposed: this confirms that we are not able to distinguish between different low-density models through GW analysis alone.
The choice of the crust does, however, impact the radii of the neutron stars involved in the coalescence. To get an estimate of the variation of the radii due to the choice of the crust alone, we compute the $M$ vs $R$ 2D distributions from the posteriors, as described in section 3 (figure \ref{2D} (b)). Our prediction of the median of the radii distribution $R^{m}$ for the 30-65 crust is slightly shifted with respect to the median obtained with the original SLy crust, and in excellent agreement with the actual MCMC result (table \ref{tabone}). We then estimate the systematic error due to the crust 
as $\Delta R^- = R_{SLy}^m - R_{30-65}^m$ and $\Delta R^+ = R_{34-35}^m - R_{SLy}^m$. We find $\Delta R_1^+ = 0.1$ km, $\Delta R_1^- = 0.2$ km and $\Delta R_2^+ = 0.1$ km, $\Delta R_2^- = 0.2$ km. 
The radii of the NSs of GW170817 then become 
$R_1 = (11.9^{+1.4 + 0.1}_{-1.4 - 0.2})$ km and $R_2 = (11.9^{+1.4 + 0.1}_{-1.4 - 0.2})$ km.

\begin{figure}[h]
\centering
\subfloat[][]
{\includegraphics[width=.45\columnwidth]{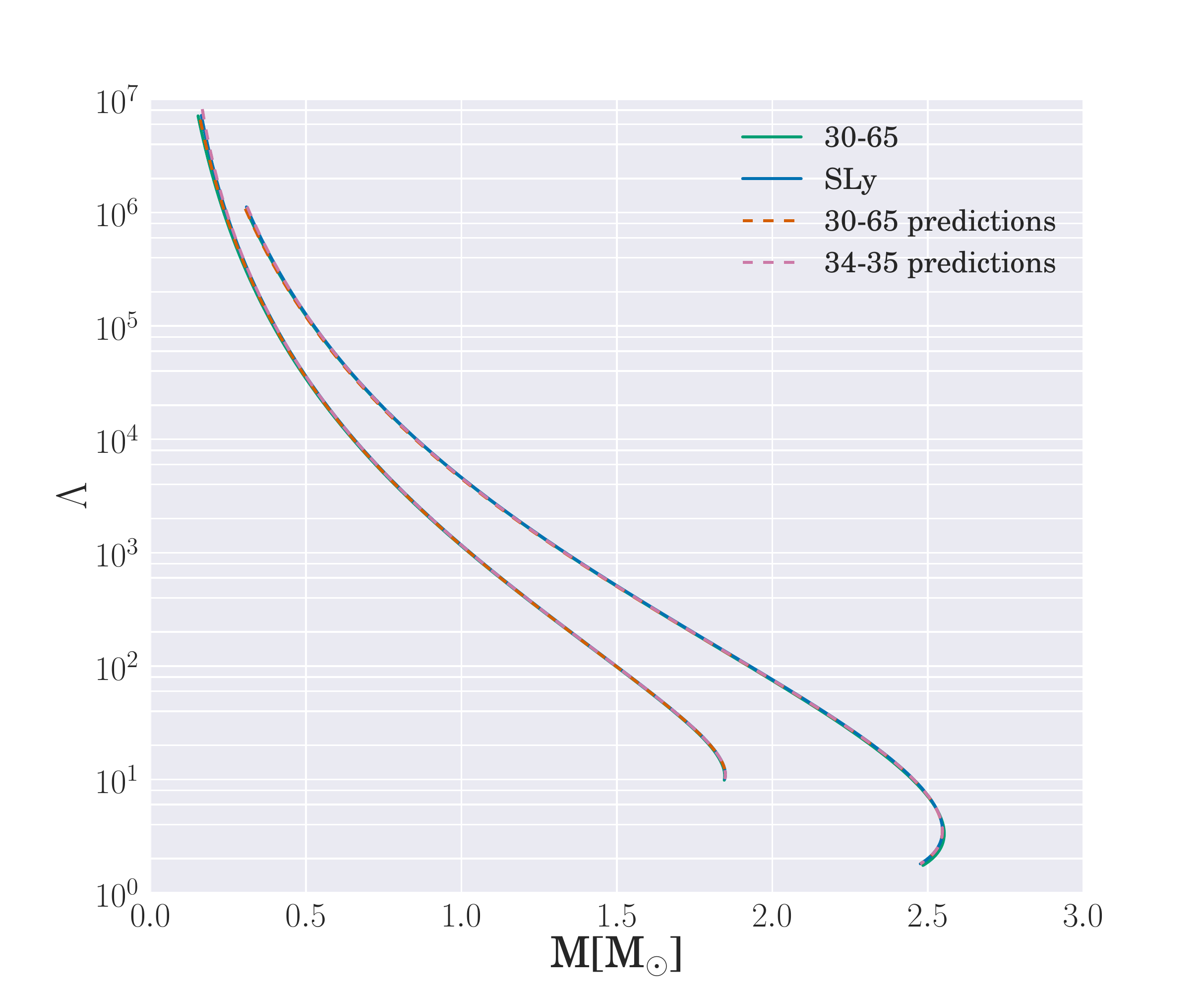}} \quad
\subfloat[][]
{\includegraphics[width=.45\columnwidth]{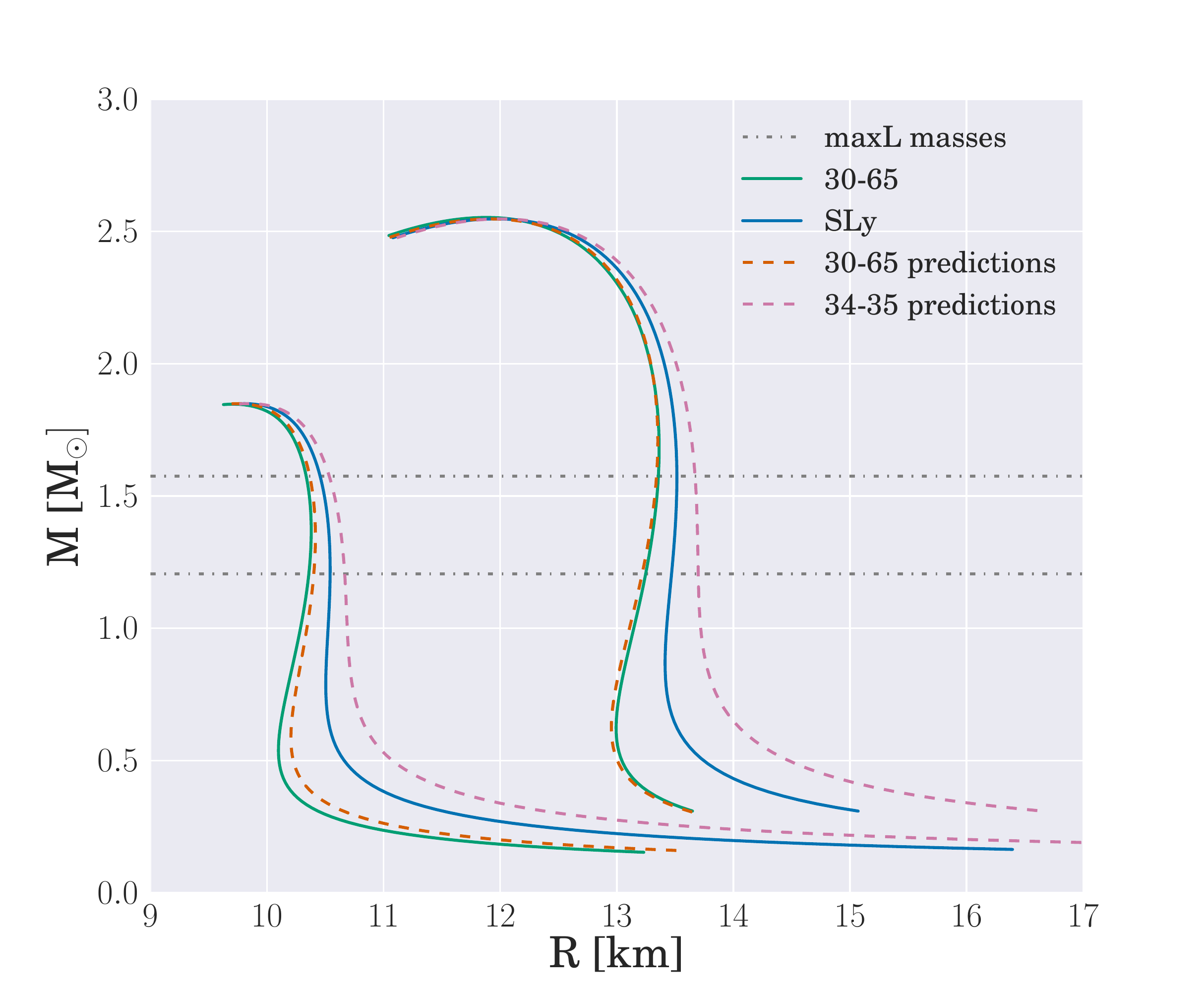}} \\
\caption{The $\Lambda(M)$ (a) and $M(R)$ (b) curves, obtained matching the crust EOSs to the 90\% CLs of figure \ref{cls}. For each percentile, $5^{th}$ or $95^{th}$, the $\Lambda(M)$ curves are on top of each other, while the $M(R)$ curves differ more as the mass decreases. This can be easily explained considering that for low-mass stars the crust has more relative importance than for heavier NS. Note also how our prediction (orange dashed) gives a good approximation of the curve obtained with the actual run (green)}
\label{curves}
\end{figure}

More physical insight can be obtained by mapping the $90\%$ pressure-density credible levels curves, appropriately glued to the selected crusts, into M(R) and $\Lambda(R)$ curves (figures \ref{curves} (a) and (b)). 
$\Lambda(M)$ curves are indistinguishable, as expected, and once a mass value $M$ has been fixed we can retrieve from the M(R) relations an estimate of the uncertainty on radius $\Delta R$. Defining $R^{X}_i$ as the radius value obtained by inverting the $X = 5^{th}$ or $X = 95^{th}$ M(R) percentile curve and setting $M = M_1$ or $M = M_2$, we have that $\Delta R$ = $\Delta R_0 + \Delta R^+ + \Delta R^-$, where $\Delta R_0$ = $ R^{95}_{SLy} - R^{5}_{SLy}$ is the original uncertainty, obtained when considering the SLy crust only, and $\Delta R^+ = R^{95}_{34-35} - R^{95}_{SLy} $, $\Delta R^- = R^{5}_{SLy} - R^{5}_{30-65}$ are the corrections which account for the different crusts. 
For $M_1 = 1.57 M_{\odot}$ and $M_2 = 1.20 M_{\odot}$ we find $\Delta R_1^+ = 0.2$ km, $\Delta R_1^- = 0.1$ km and $\Delta R_2^+ = 0.2$ km, $\Delta R_2^- = 0.1$ km. This estimate, while not very different from the one obtained earlier, gives a less complete picture of the situation as pressure-density credible levels (CLs) do not map directly into mass-radius CLs. Nonetheless, it does a reasonable job in the mass range of the binary components, and suggests that the corrections become larger as the mass becomes smaller. This behaviour can be easily explained: the less massive the star, the higher the contribution of the crust to the total mass, and the bigger the radii differences owed exclusively to the arbitrary choice of the outer layers.

\begin{table}[h]
\caption{\label{tabone} 
Left: Table containing the radii values obtained from the 2D M vs R distributions (figure 3 (b)). \\
Right: Table containing the $R^{X}_i$ radii values, obtained by matching different crusts to the $ X = 5^{th}$ or $ X = 95^{th}$ pressure-density percentile curves, inverting the implied M(R) relation and fixing $M$ to $M_1 =1.57 M_{\odot}$ or $M_2=1.20 M_{\odot}$ (figure 4 (b)).}

\lineup
\begin{tabular}{@{}*{7}{l}}

\ns
                 &\crule{2} & \crule{4} \cr
                 & $R_1 [km]$ & $R_2 [km]$ & $R_1^{5} [km]$ & $R_1^{95} [km]$ & $R_2^{5} [km]$ & $R_2^{95} [km]$  \cr
\mr
SLy (32-46) run          &  $11.9_{- 1.4}^{+ 1.4}$ &  $11.9_{- 1.4}^{+ 1.4}$ &  $10.5$    &  $13.5$  & $10.5$  & $13.5$  \cr
30-65 run        &  $11.7_{- 1.4}^{+ 1.4}$ &  $11.7_{-1.3}^{+1.4}$   &  $10.4$    &  $13.3$  & $10.4$  & $13.2$  \cr
30-65 prediction &  $11.7_{- 1.3}^{+ 1.4}$ &  $11.7_{-1.3}^{+1.4}$   &  $10.4$    &  $13.3$  & $10.4$  & $13.2$  \cr
34-35 prediction &  $12.0_{- 1.4}^{+ 1.5}$ &  $12.0_{- 1.4}^{+1.5}$  &  $10.5$    &  $13.7$  & $10.7$  & $13.7$  \cr
\br
\end{tabular}
\end{table}

\section{Conclusions}

After considering a range of realistic crust EOSs, whose parameters $S_0$ and $L$ at saturation density span the intervals $[30-34]$ MeV and $[30-70]$ MeV respectively, we selected the $S_0 = 30$ MeV and $L = 65$ MeV crust model and re-analyzed the data of GW170817. 
In parallel, we successfully predicted the outcome of the re-analysis, i.e the NS radii values. Such values were then used to compute the systematic error to be added to the radii estimates of GW170817, which amounts to a total of 0.3 km, approximately $3\%$ of R. The simple methods implemented to make predictions will likely be useful to quickly quantify the impact of the crust EOS on future radius estimates obtained through GW analyses.
Finally, we find low sensitivity of tidal parameters to the EOS at lower crust densities. GW measurements give direct information on the high density EOS, independent of uncertainties in the crust. 

\section*{Acknowledgments}
The authors thank: John Douglas Veitch, Carl-Johan Haster and Sebastian Khan for their assistance with \verb|LALSuite| installation and debugging; Benjamin Lackey and Katerina Chatziioannou for writing and sharing the plotting scripts we built on; Wynn Ho and Wolfgang Tichy for precious discussions; William Newton for supplying the low-density equations of state.
The authors also thank INFN and NSF, who supported this work through the INFN-NSF/LIGO summer student exchange program, Giancarlo Trinchero, for support at the beginning of the exchange, and Alessandro Nagar, for discussions and comments on the manuscript.
Finally, the authors are grateful for computational resources provided by the LIGO Laboratory and supported by the National Science Foundation Grants PHY-0757058 and PHY-0823459.

\section*{References}
\bibliographystyle{iopart-num}
\bibliography{refs}

\end{document}